\documentclass[]{spie}  

\usepackage{amsmath,amsfonts,amssymb}
\usepackage{graphicx}
\usepackage[colorlinks=true, allcolors=blue]{hyperref}

\title{Commissioning and on-sky performance verification of iLocater}

\author[a,*]{Marshall C. Johnson}
\author[a]{Marcelo Tala Pinto}
\author[a,$\dagger$]{Jonathan Crass}
\author[b]{Brian Sands}
\author[a]{Jackson Datkuliak}
\author[a]{Sai Vidyud Senthil Nathan}
\author[c]{Justin R. Crepp}
\author[a]{Michael Engelman}
\author[a]{Daniel Pappalardo} 
\author[a]{Julia Brady}
\author[a]{Mark Derwent}
\author[a,d]{Xavier Lesley-Salda\~na}
\author[c]{Andrew Bechter}
\author[c]{Eric Bechter} 
\author[e]{Christian Schwab}
\author[f]{David Aikens}
\author[g]{Chad Bender}
\author[a]{Chris Brandon}
\author[h]{Cynthia Brooks}
\author[i,j]{Matheus J. Castro}
\author[c]{Jeffrey Chilcote}
\author[k]{Al Conrad}
\author[l]{Gregory Delo} 
\author[a]{Erin Duell}
\author[b]{Matthew Engstrom} 
\author[g,k]{Steve Ertel} 
\author[l]{Louis Fantano}
\author[l]{John Keith Feggans}
\author[l]{Hali Flores} 
\author[a]{B. Scott Gaudi}
\author[m]{David King} 
\author[e]{Ondrej Kitzler}
\author[g]{Kaitlin M. Kratter} 
\author[l]{Jeffrey Kruk}
\author[n]{Thomas Legero}
\author[o]{Stanimir Letchev} 
\author[a]{Jerry Mason}
\author[p]{Joaquin Mason}  
\author[c]{Matthew Misch} 
\author[o]{Jacob Pember}  
\author[a,d]{Richard Pogge} 
\author[i]{Andreas Quirrenbach}
\author[l]{Joshua Schlieder} 
\author[l]{Augustyn Waczynski}
\author[k]{Joseph Shields}
\author[g,k]{Daewook Kim}
\author[k]{Mark Smithwright}
\author[a]{Allie Renshaw} 
\author[a]{Austin Richie} 
\author[c]{Alexa Rizika}
\author[c]{Nandini Sadagopan} 
\author[a]{Jonathan Shover} 
\author[b]{James Smous}
\author[a]{Jayde Spiegel} 
\author[j]{Julian Stürmer}
\author[a]{Sahil Sura}
\author[l]{Robert Switzer} 
\author[l]{Joe Thomes}
\author[a]{Ji Wang}
\author[c]{Lauren Weiss}
\author[q]{Daniel Wilson}
\author[e]{Dane Zielinski-Nicolson}

\affil[a]{Department of Astronomy, The Ohio State University, 140 West 18$^{\mathrm{th}}$ Ave., Columbus, OH 43210, USA}
\affil[b]{Engineering and Design Core Facility, University of Notre Dame, Notre Dame, IN 46556, USA}
\affil[c]{Department of Physics \& Astronomy, University of Notre Dame, 225 Nieuwland Science Hall, Notre Dame, IN 46556, USA}
\affil[d]{Center for Cosmology and AstroParticle Physics, The Ohio State University, 191 West Woodruff Avenue, Columbus, OH 43210, USA}
\affil[e]{School of Mathematical and Physical Sciences, Macquarie University, Balaclava Road, North Ryde, NSW 2109, Australia}
\affil[f]{Savvy Optics, 35 Gilbert Hill Rd., Chester, CT 06412, USA}
\affil[g]{Department of Astronomy, University of Arizona, 933 North Cherry Avenue, Tucson, AZ 85721, USA}
\affil[h]{Department of Astronomy, University of Texas at Austin, 2515 Speedway, Austin, TX 78712, USA}
\affil[i]{Landessternwarte, Zentrum für Astronomie der Universität Heidelberg, Königstuhl 12, 69117, Heidelberg, Germany}
\affil[j]{Fakultät für Physik und Astronomie, Universität Heidelberg, Im Neuenheimer Feld 226, 69120 Heidelberg, Germany}
\affil[k]{Large Binocular Telescope Observatory, 933 North Cherry Avenue, Tucson, AZ 85721, USA}
\affil[l]{NASA Goddard Space Flight Center, Greenbelt, MD 20771, USA}
\affil[m]{Institute of Astronomy, University of Cambridge, Madingley Road, Cambridge CB3 0HA, UK}
\affil[n]{Physikalisch-Technische Bundesanstalt, Bundesallee 100, 38116, Braunschweig, Germany}
\affil[o]{Max Planck Institute for Astronomy, Königstuhl 17, 69117, Heidelberg, Germany}
\affil[p]{Fathom Imaging Systems, 30 Nashua Street, Woburn, MA 01801, USA}
\affil[q]{NASA Jet Propulsion Laboratory, California Institute of Technology, 4800 Oak Grove Drive, Pasadena, CA 91109, USA}

\authorinfo{*johnson.7240@osu.edu; $\dagger$ crass.7@osu.edu}

\begin{document} 
\maketitle

\begin{abstract}
iLocater is a high-resolution near-infrared extreme precision radial velocity (EPRV) spectrograph that was deployed to the Large Binocular Telescope (LBT) in June 2026. iLocater operates over $\lambda=966$-1312 nm with a median resolving power of $R=205,000$ as measured in the laboratory. We present the commissioning and initial on-sky verification program using solar and night-time observations at the LBT. First light was achieved on 27 June 2026, and nearly 150 on-sky spectra have now been recorded. Observations include single stars ranging in spectral type from B5 to M6. iLocater uses the LBT AO system, and we have demonstrated its ability to obtain spatially resolved spectra of close ($\theta < 1''$) binary stars. 
\end{abstract}


\section{INTRODUCTION}
\label{sec:intro}  

The past decades have seen significant progress in the measurement of radial velocities (RVs) of stars using high-resolution spectrographs, which are used to discover and characterize orbiting exoplanets. 
RV measurement precision has improved from tens of m s$^{-1}$ in the 1980s to sub-m s$^{-1}$ today. The modern generation of extreme precision RV (EPRV) spectrographs are 
temperature-stabilized, and typically in vacuum, in order to minimize instrumental shifts of the spectrum which could manifest as apparent RV shifts (\citenum{Crepp14}). 
Simultaneous calibration spectra and ultra-stable references like laser frequency combs (LFCs) have allowed these spectrographs to achieve instrumental precisions which are in principle well under 1 m s$^{-1}$. The limiting factor on RV precision is now astrophysical noise due to a variety of processes in the stellar atmospheres, typically on the scale of m s$^{-1}$. See \citenum{EPRV-ARAA} for a review of the current state of the field.

One way forward for EPRV science 
to overcome this astrophysical noise barrier is to achieve high enough spectral resolutions to spectroscopically resolve stellar absorption line profile perturbations due to stellar activity and distinguish these from true Doppler shifts of the stellar spectrum due to an orbiting planet.
This requires high signal-to-noise ratio (SNR) data, which, together with high resolution, demands large collecting area. This is the central scientific goal and requirements driving the design of the iLocater spectrograph (\citenum{SPIEIntro,FinalDesign}). iLocater has been installed on the Large Binocular Telescope (LBT), the largest collecting area optical telescope currently in operation.
iLocater provides high spectral resolution in a compact instrument by operating at the diffraction limit behind the LBT's adaptive optics (AO) system, which also promotes thermal stability of the instrument. iLocater has a bandpass of $\lambda=966$-1312 nm ($Y$ and $J$ bands), red enough to allow the LBT's AO system to provide correction while blue enough to minimize thermal backgrounds and give access to a band with limited telluric contamination. The system consists of an acquisition camera (\citenum{AcCam}), cryogenic spectrograph operating at 80-100 K and stabilized to mK precision (\citenum{CryostatThermal}), and a calibration unit (\citenum{SPIE2026:calibration}). 

The acquisition camera was delivered to the LBT in 2019 (\citenum{Bechter20}). The spectrograph and calibration unit were delivered to the observatory in June 2026, and saw first light through the LBT on 27 June 2026 local (28 June UT). This followed a laboratory test campaign at The Ohio State University (OSU) to characterize the instrument (\citenum{SPIE2026:laboratory}), including some solar observations. A Menlo Systems LFC was delivered to the LBT in May 2026, which will be used to calibrate both iLocater and the PEPSI high-resolution optical spectrograph (\citenum{PEPSI}); first light has also been achieved with the LFC with both instruments. 

The June 2026 first light run included observations over the span of four nights. In this Proceedings we describe these observations and present a first look at these data, as well as the solar data gathered during the instrument laboratory campaign. We discuss plans for further commissioning observations in the fall of 2026 in order to fully demonstrate iLocater's RV capabilities.

\section{INSTRUMENT OVERVIEW}

While the overall design of the iLocater system is described in more detail in (\citenum{FinalDesign}), we describe here the aspects of the hardware and software relevant for the present discussion of the first light data.

The iLocater acquisition camera (\citenum{AcCam}) operates downstream of the LBT AO system, within the LBT Interferometer (LBTI; \citenum{LBTI}). It injects light into a 6 $\mu$m single-mode fiber (SMF) with a diameter of 50 mas on-sky (\citenum{Bechter20}).

The iLocater spectrograph is illuminated via three input SMFs imaged onto a 4096$\times$4096-pixel H4RG-10 detector from the NASA Nancy Grace Roman Space Telescope program (\citenum{H4RG,RomanH4RGflight}). During typical night-time observations, one of these (the top of the three traces) is connected to the acquisition camera on the left (SX) side of the LBT. The center fiber can serve as a simultaneous calibration reference. The bottom fiber will eventually be connected to a second acquisition camera on the right (DX) side of the telescope; however, currently it serves as a second simultaneous calibration fiber. 

When obtaining calibration data, any of the fibers may be illuminated by any calibration source (i.e., LFC, halogen flat-field lamp, uranium-neon hollow-cathode lamp, Fabry-P\'erot etalon; see \citenum{SPIE2026:calibration} for details on the calibration system). Currently, we typically obtain day-time calibration data with either the SX fiber illuminated and the others un-illuminated, or all three illuminated with the same calibration source. A full set of calibration data using all of these sources, plus dark frames, are typically taken on a daily basis when the spectrograph is cold. Due to the spectrograph optical design there is a $\sim$100-pixel offset in the dispersion direction (x-position on the detector) between the position of a given wavelength in adjacent traces of the same spectral order (which are separated from each other by 80-130 pixels in the cross-dispersion or y direction). This makes it non-trivial to combine the light from different traces as the dispersion varies across the length of each trace. The preliminary analysis presented here focuses on the analysis of only a single trace at a time. The as-built system has a median resolving power of $R=205,000$. See \citenum{SPIE2026:laboratory} for more details on the characterization of the spectrograph.

The detector reads up the ramp and produces raw data in the form of datacubes of each non-destructive detector read. 
We process these data into two-dimensional frames suitable for further analysis using a modified version of the HxRGproc software (\citenum{Ninan}). HxRGproc performs the up-the-ramp sampling to reduce the impact of read noise, and also uses the reference pixels at the detector edges to correct for the 1/f pattern noise in the detector. The modifications that we have made to HxRGproc were principally to allow the iLocater raw data format to be read into the code, mask bad pixels using a bad pixel mask, properly populate the FITS header keywords, and save data in a slightly modified format. 

For the current preliminary analysis, the two-dimensional spectra from HxRGproc are extracted into 1D using a modified version of the iLocater data simulator code (\citenum{Simulator}). This MATLAB code finds the traces using a 2D spectrum illuminated with the halogen flat-field lamp and performs a box extraction around these traces. A preliminary wavelength solution was compiled using manual identification of U and Ne lines from the atlas of \citenum{Redman12} in the uranium-neon hollow-cathode lamp spectra and a quadratic fit to the resulting line positions. This simple fit has serious limitations, particularly near the ends of the orders where the quadratic assumption breaks down, but is good to approximately 0.05 nm and is sufficient for the current purposes including assessing the bandpass (\citenum{SPIE2026:laboratory}) and identification of major stellar spectral features. Continuum normalization is performed via a simple polynomial fit, from which there is again much room for improvement. In the long run, iLocater data reduction will be performed using HxRGproc and an adapted version of the APERO pipeline (\citenum{APERO}) to address the listed shortcomings.

\section{SOLAR OBSERVATIONS}
\label{sec:solar}

On 30 January 2026 UT we obtained a set of solar observations from the laboratory at OSU. These data were obtained using an engineering-grade H4RG-10 detector.  
We set up an approximately 2.5 cm aperture focusing optic on a tracking mount pointing out the laboratory window. This optic focused sunlight into an SMF which was then routed into the spectrograph, illuminating both the top and bottom traces. This analysis focuses on the top fiber only. 
We obtained 54 individual exposures over the course of approximately one hour during the afternoon. During this time the airmass increased from approximately 2.5 to 4. 

\begin{figure} [ht]
   \begin{center}
   \begin{tabular}{c} 
   \includegraphics[width=\textwidth]{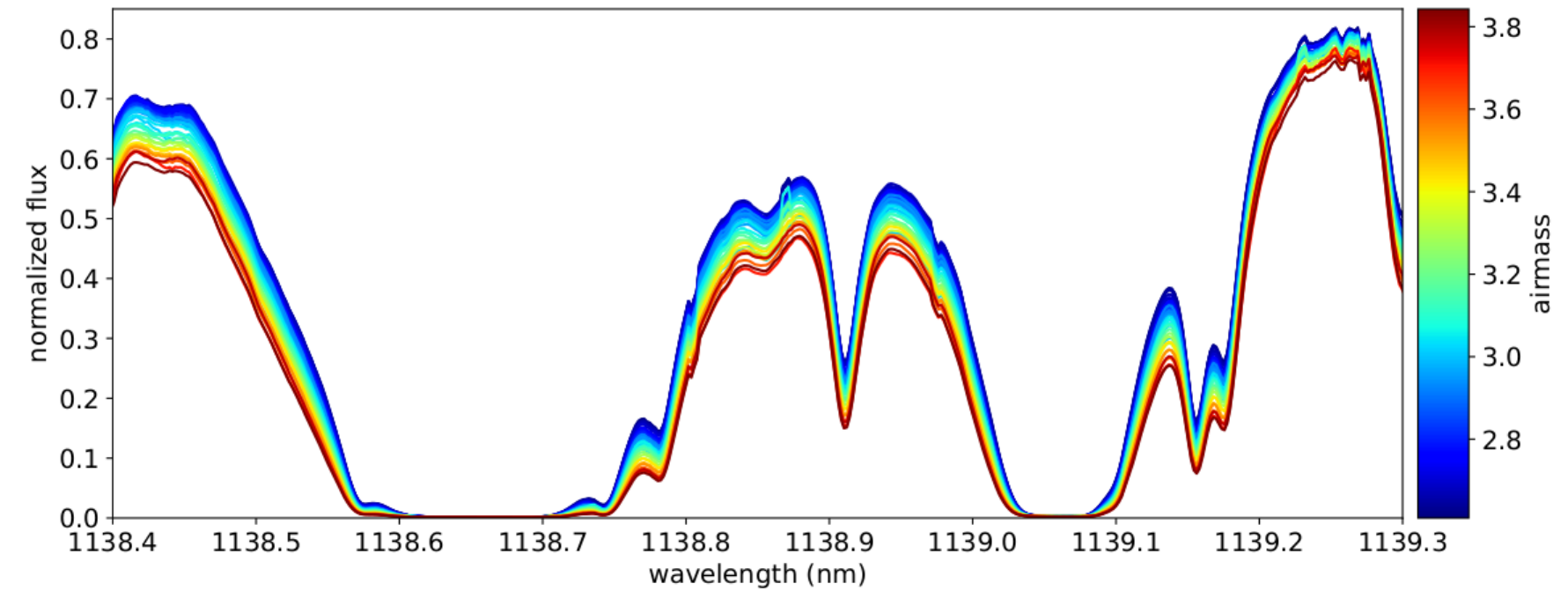}
   \end{tabular}
   \end{center}
   \caption[example] 
   { \label{fig:solar_tellurics} 
A small portion of one iLocater order in solar data obtained through the lab window at OSU on 30 January 2026, in the region of strong tellurics between the $Y$ and $J$ bands. Successive spectra in these time-series observations are color-coded by airmass, which increased from approximately 2.5 to 4 over the one-hour course of the observations. The damping wings of these saturated lines can be seen broadening as the absorbing column increases.}
   \end{figure}

In Fig.~\ref{fig:solar_tellurics} we show a section of the spectra in the strong telluric region between the $Y$ and $J$ bands. The damping wings of these strong lines clearly broaden as the airmass and absorbing column increases, showcasing the potential of iLocater for studying telluric absorption at high spectral resolution.

We also obtained 10 additional solar spectra at the LBT during the first light run between 24 and 27 June 2026. These illuminated all three fibers using a feed from the the PEPSI solar telescope (\citenum{PEPSISolar}). Data analysis is ongoing. 

\begin{figure} [ht!]
   \begin{center}
   \begin{tabular}{c} 
   \includegraphics[trim=0 0 0.1cm 0, clip=true, width=0.98\textwidth]{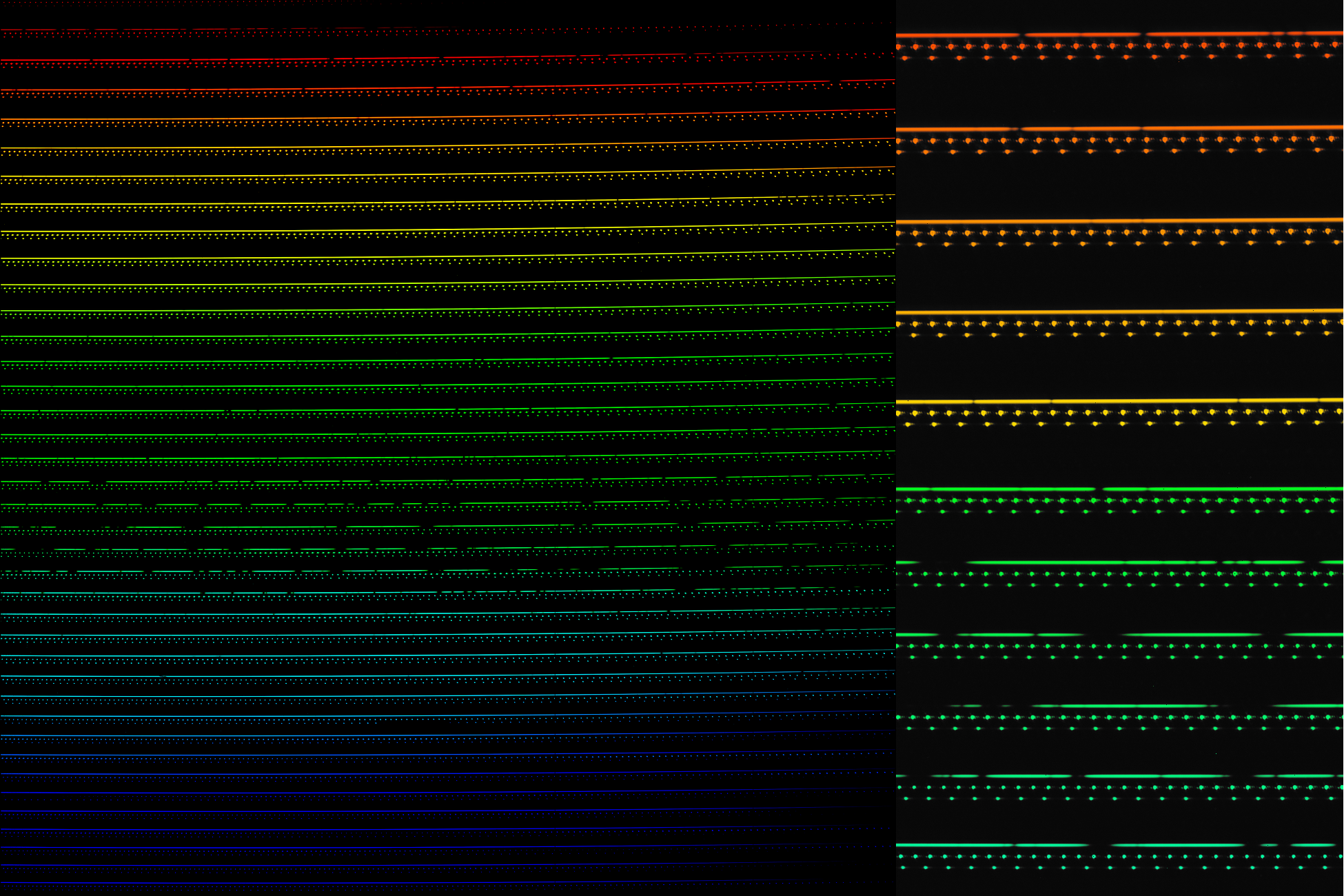}
   \end{tabular}
   \end{center}
   \caption[example] 
   { \label{fig:vega} 
An example iLocater spectrum of our first light target, Vega. The left panel shows the full 4096$\times$4096 pixel detector, while the two right panels show cut-outs of two representative regions at full resolution. The top trace is illuminated with Vega, the middle trace with the LFC, and the bottom with the Fabry-P\'erot etalon. The image has been artificially colored to highlight the bandpass; overall wavelengths increase from the blue end at the bottom of the detector to the red end of the bandpass at the top, while within each individual order the wavelengths increase from left to right. The $J$ band is at top and $Y$ is at bottom, separated by a region of strong telluric absorption. A 1D wavelength-calibrated version of this spectrum is shown in Fig.~\ref{fig:spectral_sequence} along with other targets.}
   \end{figure}

\section{FIRST LIGHT OBSERVATIONS}
\label{sec:firstlight}

iLocater achieved first light through the LBT on the night of 27 June 2026 (28 June UT at approximately 04:47) with a spectrum of Vega ($\alpha$ Lyr). An echellogram of Vega is shown in Fig.~\ref{fig:vega}. Unlike the solar data described in \S\ref{sec:solar}, these data were obtained using a science-grade H4RG-10 detector.

Over the course of four nights (27-30 June 2026, or 28 June-1 July UT), we gathered a total of 142 on-sky exposures of 31 individual targets. The overall objectives of these observations were to:

\begin{enumerate}
    \item Inject light from the LBT into the spectrograph via the acquisition camera.

    \item Observe a series of targets with different spectral types and magnitudes in order to assess the efficiency of the overall iLocater optical system, including the interaction of iLocater's newly installed hardware with light delivered by the LBT's AO system. 

    \item Assess the iLocater system performance on close binary targets by conducting two separate tests. First, by observing systems consisting of an early-type and a late-type component; the early-type star should lack sharp spectral lines, and so any sharp non-telluric lines observed are contamination due to the other star. Second, by centering a bright, single star target and stepping the fiber off the target in order to record how much residual light is present as a function of separation.

    \item Begin building a library of spectra of hot, rapidly rotating telluric standard stars that will be used for APERO's telluric correction routines (\citenum{APERO}).

    \item Obtain on-sky spectra with the calibration fibers illuminated to assess any stray light contamination in the science trace.
\end{enumerate}

We obtained data to address all of these objectives, and data analysis is ongoing. Here, we highlight a few initial results.

\subsection{Spectral Sequence}

\begin{figure}
   \begin{center}
   \begin{tabular}{c} 
   \includegraphics[trim=0 0.5cm 0 0.5cm, clip=true, width=\textwidth]{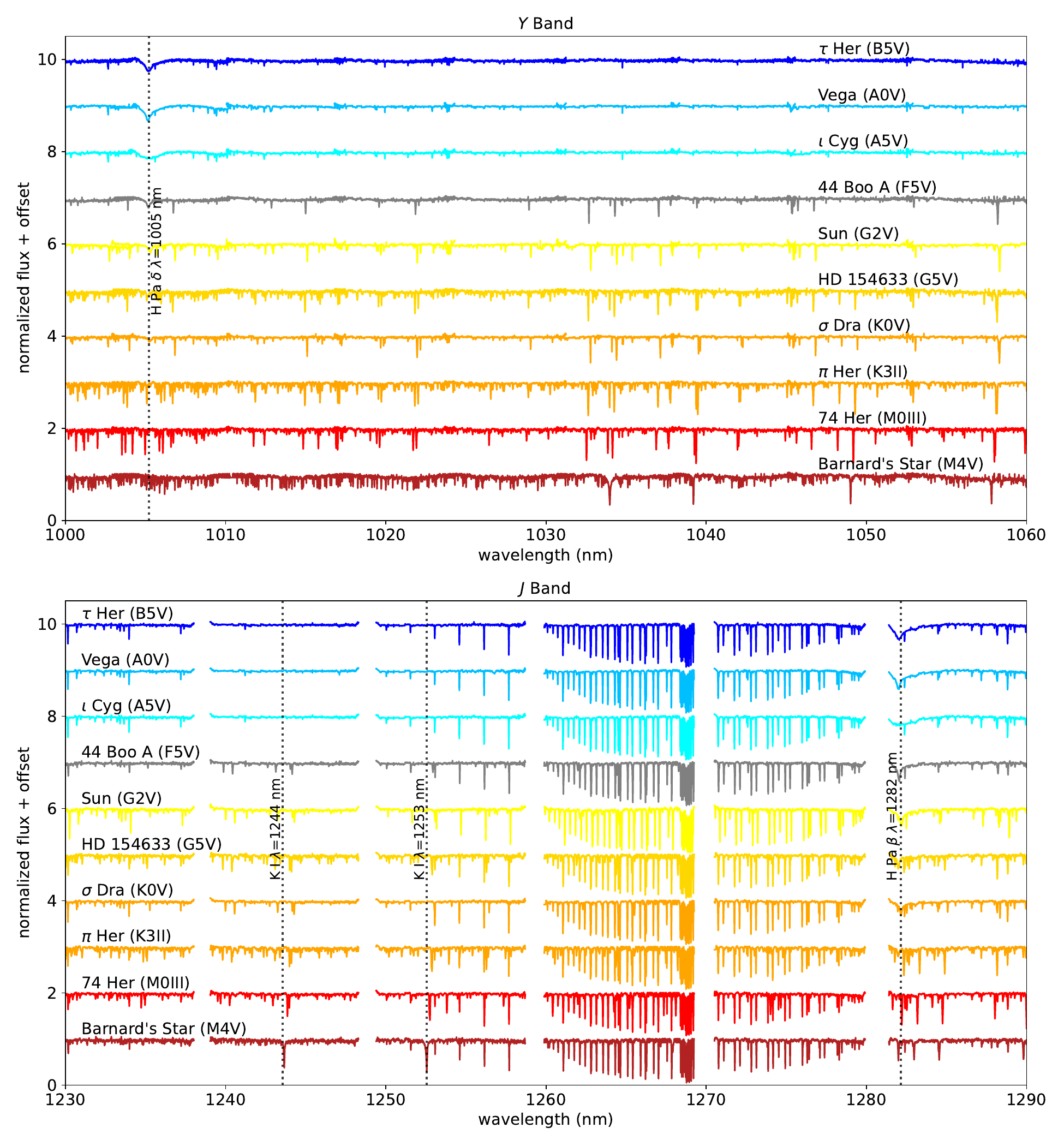}
   \end{tabular}
   \end{center}
   \caption[example] 
   { \label{fig:spectral_sequence} 
A spectral sequence showing targets observed by iLocater during the first light run, with effective temperature $T_{\mathrm{eff}}$ increasing from bottom to top. Each spectrum is offset vertically for clarity and is labeled with the name of the star and the spectral type. The top and bottom panels show a portion of iLocater's spectral bandpass within the $Y$ and $J$ bands, respectively. The wavelength coverage is continuous in the $Y$ band but there are inter-order gaps in the $J$ band. Several astrophysical absorption lines lines of interest are labeled: the H I Paschen $\beta$ and $\delta$ lines, which are strongest in the hottest stars, and the K I doublet at $\lambda\lambda=1244$, 1253 nm, which are strongest in the coolest stars. The forests of sharp lines in the $J$ band which are common between all stars are telluric absorption due to molecules (mostly H$_2$O) in the Earth's atmosphere, while for the hotter stars iLocater's resolution allows the rotationally-broadened stellar lines to be distinguished from tellurics by eye. The continuum normalization and wavelength solution are works in progress. Spectra are shown in the observatory rest frame.}
   \end{figure}

In Fig.~\ref{fig:spectral_sequence} we highlight several of the targets observed with iLocater, forming a spectral sequence ranging from B5 to M4 type, and including both dwarf and giant stars. These targets range in $R$ ($J$)-band magnitude from 0.0 to 8.3 (0.0 to 5.2) Analysis of these data will allow us to assess the SNR as a function of exposure time and spectral type, allowing a measurement of the overall instrument efficiency; this is critical for planning future instrument observations. 

\subsection{AO Performance}

\begin{figure}
   \begin{center}
   \begin{tabular}{c} 
   \includegraphics[width=0.7\textwidth]{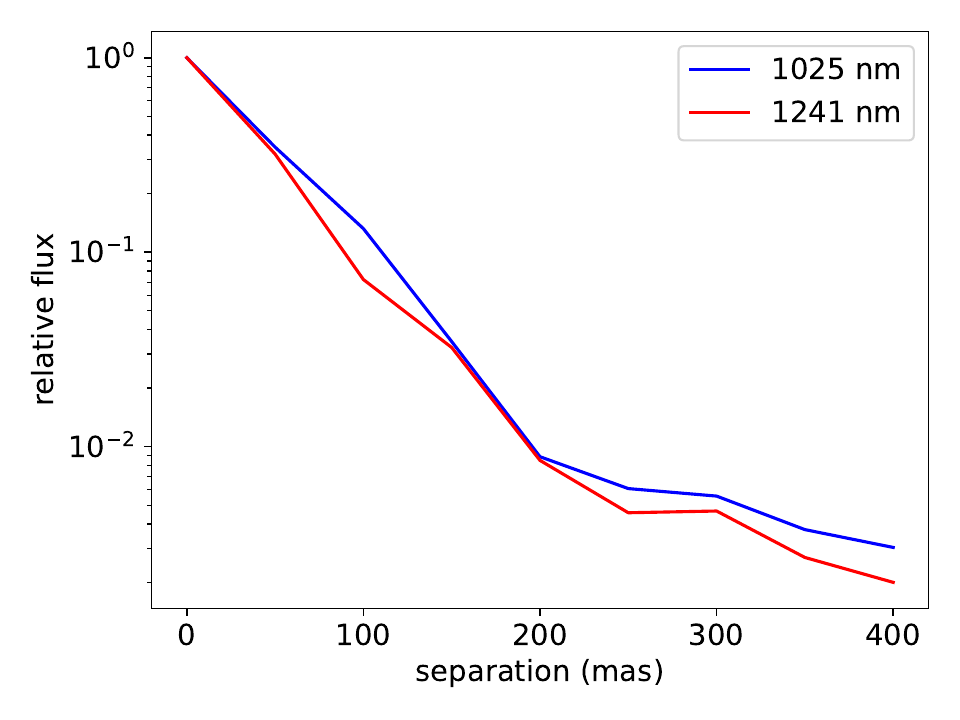}
   \end{tabular}
   \end{center}
   \caption[example] 
   { \label{fig:contrast_curve} A preliminary pseudo-``contrast curve'' obtained with iLocater observations of Deneb to test the performance of the spectrograph behind the LBT AO system and the impact of contamination upon observations of close binary stars. Each line shows the mean flux recovered by the spectrograph relative to that of the on-axis reference spectrum in two narrow spectral regions near wavelengths of 1025 (blue) and 1241 (red) nm as a function of the fiber separation from the central target. Analysis is preliminary. 
}
   \end{figure}

In order to assess iLocater's performance with the LBT AO system and the potential impact of contaminating light in observations of close binaries, we observed the bright star Deneb ($\alpha$ Cyg). After collecting an on-axis reference spectrum, we stepped the fiber off of the target in successive increments of approximately 50 mas and at each position recorded a spectrum with increasing exposure times (ranging from 14 seconds to 7 minutes). In Fig.~\ref{fig:contrast_curve} we show a preliminary analysis of these data. For each spectrum, we compute the mean flux (up-the-ramp slope) measured by HxRGproc and extracted using the MATLAB pipeline in two narrow spectral regions near 1025 and 1241 nm, chosen to be near the peak of the blaze function and free of strong telluric or stellar features. The recovered flux decreases with increasing separation, dropping by a factor of $>100$ outside of approximately 200 mas. This is consistent with AO residuals noted in imaging in the acquisition camera (e.g., \citenum{AcCam,2Cyg}).

We also obtained spectra of several close binary systems (separations $<1$ arcsec) in order to directly test for spectral contamination; analysis of these data is ongoing.

\subsection{Simultaneous PEPSI Spectroscopy}

iLocater currently has only a single acquisition camera, on the left (SX) side of the LBT, allowing the other (DX)
primary to be available for simultaneous observations with other instruments. During the first light run, we successfully demonstrated simultaneous high-resolution optical spectroscopy with the PEPSI spectrograph (\citenum{PEPSI}), covering $\lambda=383.7$-906.7 nm in three wavelength settings. We used the PEPSI medium ($R=130,000$) or high ($R=270,000$) resolution mode depending upon the target brightness. Successful demonstration of this joint instrument mode allows iLocater+PEPSI users to obtain high-resolution spectra covering most of the optical and near-infrared simultaneously. 

Other instruments which may be used simultaneously with iLocater, depending upon the science case, include the SHARK-VIS high-resolution imager (\citenum{SHARK-Vis}), the LUCI2 infrared imager and spectrograph (\citenum{LUCI}), the LBC-Red large-format imager (\citenum{LBC-red}), and the MODS2 multi-object spectrograph (\citenum{MODS}).

\section{THE ILOCATER COMMISSIONING PLAN}

The first-light observations described in \S\ref{sec:firstlight} are merely the first part of iLocater commissioning. During these observations the spectrograph was still thermally equilibrating, limiting our ability to fully assess the instrument's RV stability. During the summer of 2026, we will obtain a series of daily calibration data in order to begin assessing iLocater's RV performance at a stable operating temperature. 

Night-time commissioning observations will continue during the the 2026B LBT observing semester, covering approximately September 2026-January 2027. These will focus on the RV performance of the instrument. We will obtain observations of a set of RV standard stars, drawn from the NEID RV standard list and likely to be coordinated with NEID. These will provide a baseline of expected near-zero RV variability, and the NEID spectra (as well as simultaneous PEPSI spectroscopy) will provide a reference to assess iLocater's RV performance.

Solar observations will also continue to be part of iLocater's program of standard star observations and an important part of the verification of the instrument's RV performance. These will provide a near-infrared, very high-resolution view of solar activity and sun-as-a-star RV data, which has been key to recent progress in data analysis methods to address the problem of stellar activity in RVs (e.g., \citenum{Dumusque26}).
Together with existing solar feeds at optical EPRV spectrographs including HARPS-N (\citenum{HARPS-Nsolar}), NEID (\citenum{NEIDsolar}), EXPRES (\citenum{EXPRESsolar}), and KPF (\citenum{KPFsolar}), and a near-infrared feed to NIRPS (\citenum{NIRPS}), as well as the existing feed from the same solar telescope to the LBT's PEPSI high-resolution optical spectrograph (\citenum{PEPSISolar}), the iLocater solar dataset will open a new wavelength/resolution regime for sun-as-a-star studies and allow comparison to simultaneous optical data from instruments at similar longitudes like NEID and EXPRES.

Additionally, we will observe a handful of known hot Jupiter hosts in order to demonstrate that we can recover known astrophysical RV signals. Hot Jupiters are optimal targets as they have short-period orbits (enabling all orbital phases to be covered within a few-night observing run) and RV semi-amplitudes of tens to hundreds of m s$^{-1}$, which should easily be recoverable with an EPRV spectrograph like iLocater.

Finally, we will observe exoplanetary transits. The Rossiter-McLaughlin (R-M) effect during the transit provides a known RV signal over the course of a few hours, which can be observed in a single night to assess and demonstrate iLocater's short-term RV stability. We have assembled a target list of planets with published R-M observations to which we can compare, and measured R-M amplitudes of at least 20 m s$^{-1}$ in order to ensure easy detectability. A second transit observation will demonstrate that iLocater is ready to pursue early science by observing a planet with known He~\textsc{i} metastable triplet absorption. These lines, located at 1083 nm, are within the iLocater bandpass and have emerged in recent years as a key tracer of escaping atmospheres from exoplanets (e.g., \citenum{OklopcicHirata}). Detecting this known absorption signal will demonstrate that iLocater is ready to pursue early science observations.

Additional work and development will include finishing modifications of the APERO pipeline (\citenum{APERO}) to handle iLocater data and measure RVs; adjustments to the spectrograph optical system to optimize the bandpass; and work to use the new LBT LFC to provide a stable and accurate wavelength solution.

\section{CONCLUSIONS}

iLocater, the new near-infrared EPRV spectrograph for the LBT, was delivered to the observatory, re-integrated, and saw first light in June 2026. Analysis of the nearly 150 spectra obtained during the four-night first light run is ongoing. Further commissioning observations during the fall of 2026 will demonstrate iLocater's RV capabilities and that the instrument is ready to pursue shared-risk early science. 

\acknowledgments 

This material is based upon work supported by the National Science Foundation under Grant Nos. 1654125, 2108603 and 2408424, the National Aeronautics and Space Administration under Space Act Agreement No. 38232 and Grant No. 80NSSC24K1444, and The Mt. Cuba Astronomical Foundation. J.R.C. acknowledges partial support for the iLocater project from the Wolfe family and Potenziani family.

The LBT is an international collaboration among institutions in the United States and Europe. At the time data were acquired for this research, LBT Corporation Members were the The Ohio State University, representing The Ohio State University, University of Notre Dame, University of Minnesota, and University of Virginia; the University of Arizona on behalf of the Arizona Board of Regents; and Istituto Nazionale di Astrofisica, Italy.  This research used the facilities of the Italian Center for Astronomical Archives (IA2) operated by INAF at the Astronomical Observatory of Trieste.  Observations have benefited from the use of ALTA Center (alta.arcetri.inaf.it) forecasts performed with the Astro-Meso-Nh model. Initialization data of the ALTA automatic forecast system come from the General Circulation Model (HRES) of the European Centre for Medium Range Weather Forecasts. iLocater takes advantage of the common infrastructure provided by the Large Binocular Telescope Interferometer (LBTI).

\bibliography{report} 
\bibliographystyle{spiebib} 

\end{document}